\documentclass[12pt]{iopart}
\usepackage{amsfonts}
\expandafter\let\csname equation*\endcsname\relax
\expandafter\let\csname endequation*\endcsname\relax
\usepackage{amsmath}
\usepackage{amssymb}
\usepackage{graphicx}
\usepackage{bm}
\usepackage{hyperref}
\usepackage[usenames]{color}
\newcommand{\prl}{Phys. Rev. Lett.}
\newcommand{\pra}{Phys. Rev. A}
\newcommand{\nt}{Nature}
\newcommand{\sci}{Science}
\newcommand{\lp}{Laser Phys.}
\begin{document}
\title{Raman superradiance and spin lattice of ultracold atoms in optical cavities}
\author{S Safaei$^1$,\"O E M\"ustecapl{\i}o\u{g}lu$^2$ and B Tanatar$^1$}
\ead{safaei@fen.bilkent.edu.tr}
\address{$^1$ Department of Physics, Bilkent University, Bilkent, 06800 Ankara, Turkey}
\address{$^2$ Department of Physics, Ko\c{c} University, Sar{\i}yer, 34450 Istanbul, Turkey}
\begin{abstract}
We investigate synthesis of a hyperfine spin lattice in an atomic Bose-Einstein 
condensate, with two hyperfine spin components, inside a one-dimensional high-finesse 
optical cavity, using off-resonant superradiant Raman scattering. Spatio-temporal 
evolution of the relative population of the hyperfine spin modes is examined numerically 
by solving the coupled cavity-condensate mean field equations in the dispersive regime. 
We find, analytically and numerically, that beyond a certain threshold of the transverse 
laser pump, Raman superradiance and self-organization of the hyperfine spin components 
simultaneously occur and as a result a magnetic lattice is formed. The effects of an extra 
laser pump parallel to the cavity axis and the time-dependence of the pump strength 
on the synthesis of a sharper lattice are also addressed.
\end{abstract}
\submitto{New Journal of Physics}
\maketitle
\section{Introduction}
\label{intro}
An atomic gas inside a high-finesse optical cavity \cite{mekhov_quantum_2012,ritsch_cold_2013} 
may exhibit self-organization when it is subjected to a transverse laser pump 
\cite{domokos_collective_2002,black_observation_2003,
nagy_self-organization_2008,konya_multimode_2011,bux_cavity-controlled_2011}. In matter-cavity Quantum 
Electrodynamics (QED) systems, the mechanical effect of the electromagnetic fields 
on the motional states of atoms and phase shift effect of atomic motion on the fields 
induce each other mutually in a self-consistent loop. 
The idea of trapping atomic Bose-Einstein condensates (BEC) in high-finesse 
optical cavities \cite{horak_coherent_2000} has been experimentally realized and developed 
\cite{brennecke_cavity_2007,colombe_strong_2007,slama_superradiant_2007,wolke_cavity_2012} 
to allow for sufficiently strong cavity-condensate coupling in order to 
realize nonlinear effects, such as bistability, even with cavity photon number below unity 
\cite{elsasser_optical_2004,gupta_cavity_2007,ritter_dynamical_2009,safaei_bistable_2013} 
and to probe quantum phases of the condensate by cavity photons \cite{mekhov_probing_2007}. 
Quite recently, Dicke superradiance quantum phase transition  
\cite{dicke_coherence_1954,hepp_superradiant_1973,wang_phase_1973} in a BEC-cavity 
system \cite{nagy_dicke-model_2010} has been demonstrated \cite{baumann_dicke_2010,baumann_exploring_2011} 
and the nonequilibrium dynamics of such systems have been studied \cite{bhaseen_dynamics_2012,torre_keldysh_2013} 
taking into account the finite size effects \cite{konya_finite_2012,oztop_excitations_2012} 
and examining nonequilibrium effects at the critical point \cite{oztop_excitations_2012,nagy_critical_2011}. 
 
Dicke quantum phase transition for the single-mode BEC inside a high-finesse cavity \cite{baumann_dicke_2010} 
is characterized by an abrupt increase in the number of cavity photons, after a certain 
threshold of the pump intensity, which is accompanied by broken translational symmetry 
of the condensate with the formation of an optical lattice \cite{yukalov_cold_2009}.
Pump-cavity photon scattering couples the initial zero-momentum state of BEC to a 
superposition of higher recoil momentum states \cite{bhaseen_dynamics_2012}. 
A quite different scenario happens if a condensate of atoms with two different hyperfine 
states is pumped by a laser field far-detuned from the atomic transition \cite{schneble_raman_2004,
yoshikawa_observation_2004}, that is Raman superradiance \cite{cola_theory_2004,uys_theory_2007}  
may occur during which the hyperfine state of atoms changes. 

There has been much interest in the multi-mode atom-cavity systems recently, such as bosonic 
Josephson junctions inside a single-mode cavity \cite{huang_cavity-induced_2011} and spin 
glasses of single-component BEC in a multi-mode cavity \cite{strack_dicke_2011}. Optical 
bistability has been studied in spin-1 \cite{zhou_cavity-mediated_2009,zhou_spin_2010} 
and in two-mode BECs \cite{safaei_bistable_2013,dong_multistability_2011}. Multi-species 
systems provide a very rich platform for investigation of phase transitions, in addition 
to their practical advantages such as faster self-organization with lower threshold 
\cite{grieser_selforganisation_2011}, and efficient, easily interpretable imaging of correlations in 
phase transition by the cavity field \cite{guo_cavity-enhanced_2009,oztop_excitations_2012}.

In this article, we examine the idea of the Dicke-like phase transition in a system of 
BEC-cavity with Raman coupling as well as formation of magnetic lattices in the condensate. 
We consider a two-mode BEC, where the two modes correspond to two hyperfine states of the 
atoms, inside a one-dimensional optical cavity pumped by a laser field perpendicular to the 
cavity axis. The two modes of the condensate are coupled in a Raman scheme through cavity mode 
and laser field. The laser field and the cavity mode are far detuned from the atomic transition 
and therefore the system is in dispersive regime. Moreover, the laser field is slightly detuned 
from the cavity resonance. Numerically solving the coupled nonlinear dynamical equations 
of the system, we show that beyond a certain value of the transverse pump strength, atoms scatter 
the laser field into the cavity mode and in return, themselves move to the higher hyperfine 
state. As a result, Raman superradiance and translational symmetry breaking of the condensate 
take place simultaneously. The latter, which is a result of self-organization of atoms in the 
higher hyperfine state, leads to formation of ferromagnetic or ferrimagnetic lattice, depending 
on the rate of Raman transition. In this work we have also addressed the effect of an extra 
parallel pump and time-dependence of the transverse pump on the synthesis of a well-defined 
magnetic lattice. 

The rest of this paper is organized as follows. 
In Sec.~\ref{model} we introduce our BEC-cavity model and derive the mean-field equations which 
govern the dynamics of the cavity-matter system. In Sec.~\ref{etac}, we calculate the critical 
value of transverse pump strength after which the Raman superradiance takes place using first 
order perturbation approach. 
Results of the numerical solution of dynamical equations are presented in Sec.~\ref{numerics}. 
In Sec.~\ref{practical} we discuss and address the effect of an extra parallel pump and 
time-dependent pump for practical synthesis of a sharp and stable magnetic lattice. We 
summarize our results in Sec.~\ref{conc}.

\begin{figure}[!t]
\begin{center}
\includegraphics[width=0.6\textwidth,angle=0]{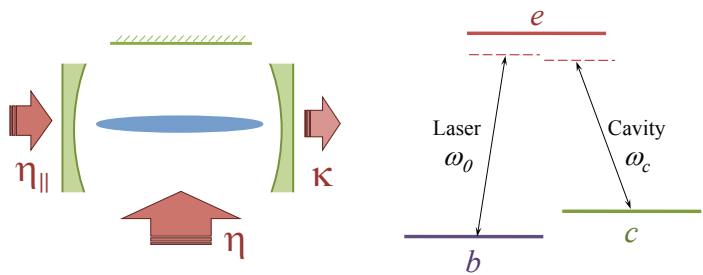}
\caption{\label{fig:cavity-raman} (Color online) 
Left: Schematic drawing of a BEC in one dimensional optical cavity subject to 
parallel and transverse laser fields. Cavity has a decay rate of $\kappa$. Right: The 
doublet of lower levels ($b$ and $c$) of the BEC atoms are coupled by cavity field and laser 
field via the atomic excited state $e$ in Raman scattering scheme. Both laser field and 
cavity field are far detuned from the atomic transition frequencies.}
\end{center}
\end{figure}
\section{Model}
\label{model}
We consider a condensate of $N$ atoms with two non-degenerate hyperfine states, 
$|b\rangle$ and $|c\rangle$, in a one-dimensional single-mode cavity of frequency 
$\omega_c$ as shown in Fig.~\ref{fig:cavity-raman}.
With an appropriate design of a trap one can isolate two desired 
hyperfine states ($|m_F=-1\rangle$ and $|m_F=1\rangle$) from 
the rest. In such a case, since other hyperfine states will be expelled from the trap, any 
inelastic atomic collision resulting in transition of atoms to other hyperfine states would 
lead to particle loss from the trap. At low temperatures the rate of particle loss is very 
small in cold atomic gases and specifically the condensates 
\cite{kagan_effect_1985,burt_coherence_1997,ketterle_coherence_1997}. 
We, thus, omit particle loss in the present calculations.
The hyperfine states are coupled to an excited state $|e\rangle$ (with $m_F=0$) 
by a transverse pump of frequency $\omega_0$ and the cavity field in the Raman scheme 
where coupling of each field to the other transition, as well as 
coupling of the other cavity mode with same frequency but opposite polarization, are 
forbidden due to conservation of angular momentum (a system in which both polarizations 
of cavity field are coupled to several hyperfine states is studied in Ref.~\cite{gangl_cavity_2002}).
The cavity is driven by another laser with the 
same frequency $\omega_0$. In the dispersive regime where the fields are far-detuned 
from the atomic transitions, the Hamiltonian of the system 
\cite{cola_theory_2004,uys_theory_2007} can be written as
\begin{eqnarray}
\label{eq:H1}
H&=&\sum_{j=b,c}\int dx~\psi^{\dag}_j\left(-\frac{\hbar^2}{2m}\frac{\partial^2}{\partial x^2}
+ V_j(x) + \hbar\omega_{bc}\delta_{j,c}\right)\psi_j+H_{\text{Raman}}\nonumber\\
&+& \sum_{i,j=b,c} \int
dx~\frac{u_{ij}}{2}\psi^{\dag}_i\psi^{\dag}_j\psi_j\psi_i
+\hbar\omega_c a^{\dag}a-i\hbar\eta_{||}(ae^{i\omega_0t}-a^{\dag}e^{-i\omega_0t}),
\end{eqnarray}
where $\psi_j(x,t)$ $(\psi_j^\dag(x,t)$) is the annihilation (creation) operator for a 
bosonic atom at space-time point $(x,t)$. $V_j(x)$ is the external trap potential for 
the state $j=b,c$ and $\omega_{bc}$ is the frequency of $b\leftrightarrow c$ transition. 
$\{u_{ij}\}$ are the interaction strengths of atoms in states $i$ and $j$ and are related 
to $s$-wave scattering lengths $\{a_{ij}\}$ through $u_{ij}=4\pi\hbar^2a_{ij}/(mw^2)$ with 
$m$ being the mass of atoms and $w$ is the transverse size of the condensate. The parallel 
laser field strength is denoted by $\eta_{||}$ and the annihilation (creation) operator of 
the cavity mode is $a$ ($a^{\dag}$). For a cavity mode with wave number $k$, Raman scattering 
Hamiltonian ($H_{\text{Raman}}$) has the form
\begin{eqnarray}
\label{eq:Hraman1}
H_{\text{Raman}}=&-&i\hbar\int dx~\psi^{\dag}_e h_0
(e^{-i\omega_0t}+e^{i\omega_0t})
\psi_b + H.c.\nonumber\\
&-&i\hbar\int dx~\psi^{\dag}_e g_0\cos(kx)(a+a^{\dag})\psi_c + H.c.,
\end{eqnarray}
where $h_0$ and $g_0$ are the atom-pump and atom-cavity dipole interaction strengths, respectively. 
Transverse pump profile is assumed to be wide enough to take $h_0$ uniform. Dipole approximation is
used for the transverse direction. After adiabatically eliminating $\psi_e$, under the condition of 
$\Delta_0=\omega_0-\omega_{be}$ being larger than the excited state linewidth, the Hamiltonian reduces to
\begin{eqnarray}
\label{eq:H2}
H = \sum_{j=b,c}\int dx~\psi^{\dag}_j{\cal{H}}\psi_j 
&+& \sum_{i,j=b,c} \int dx~\frac{u_{ij}}{2}\psi^{\dag}_i\psi^{\dag}_j\psi_j\psi_i\nonumber\\
&-&\hbar\delta_c a^{\dag}a-i\hbar\eta_{||}(a-a^{\dag})
\end{eqnarray}
where, in a rotating frame defined by the unitary operator $U=\exp(-i\omega_0 t a^{\dag}a)$,
\begin{eqnarray}
\label{eq:Hsingle}
{\cal{H}}=&-&\frac{\hbar^2}{2m}\frac{\partial^2}{\partial x^2} 
+\left(\frac{2\hbar h_0^2}{\Delta_0}+V_b(x)\right)\sigma^-\sigma^+
\nonumber\\
&+&\left(\hbar U_0\cos^2(kx)(aa^{\dag}+a^{\dag}a)+V_c(x)+\hbar\omega_{bc}\right)\sigma^+\sigma^- 
\nonumber\\
&+&\hbar\eta(a+a^{\dag})\cos(kx)(\sigma^- + \sigma^+).
\end{eqnarray}
Here $U_0=g_0^2/\Delta_0$, $\eta=h_0g_0/\Delta_0$, $\delta_c=\omega_0-\omega_c$ \cite{brennecke_cavity_2007}, 
$\sigma^+=|c\rangle\langle b|$ and $\sigma^-=|b\rangle\langle c|$.

Early stages of the dynamics are strongly influenced by quantum fluctuations that 
trigger the superradiance. We consider the late time dynamics in which the condensate 
and field variables are assumed to be classical \cite{uys_theory_2007}. The effect of 
quantum fluctuations is introduced by seeding the cavity field in numerical simulations. 
In our case seeding is performed either by adding very small fluctuations proportional 
to $\cos(kx)$ to $\psi_c$ (c.f. Sec.\,\ref{PT}) when there is no parallel pump, or physically 
by the parallel pump that drives the cavity (c.f. Sec.\,\ref{practical}). The Heisenberg 
equations of motion in this mean-field regime take the following form
\begin{eqnarray}
\dot{\psi}_b=
&-&\frac{i}{\hbar}\left(
-\frac{\hbar^2}{2m}\frac{\partial^2}{\partial x^2}+V_b(x)
+\frac{2\hbar h_0^2}{\Delta_0}
+u_{bb}|\psi_b|^2 + u_{bc}|\psi_c|^2
\right)\psi_b\nonumber\\
&-&\frac{i}{\hbar}V_1\psi_c
\label{eq:eom-psib}\\
\dot{\psi}_c=
&-&\frac{i}{\hbar}\left(
-\frac{\hbar^2}{2m}\frac{\partial^2}{\partial x^2}
+V_c(x)+\hbar\omega_{bc}+V_2
+u_{cc}|\psi_c|^2+u_{bc}|\psi_b|^2
\right)\psi_c\nonumber\\
&-&\frac{i}{\hbar}V_1\psi_b
\label{eq:eom-psic}
\end{eqnarray}
\begin{eqnarray}
\dot{\alpha}&=&i\left(i\kappa+\delta_c-2U_0\int dx~|\psi_c|^2\cos^2(kx)\right)\alpha
\nonumber\\
&-&i\eta\int dx~\cos(kx)(\psi^*_c \psi_b + \psi^*_b\psi_c)+\eta_{||},
\label{eq:eom-alpha}
\end{eqnarray}
where $\alpha_r$ is the real part of the cavity field $\alpha$ and $\kappa$ is the phenomenological 
decay rate of the cavity  \cite{uys_theory_2007}. Here $V_1=2\hbar\eta\cos(kx)\alpha_r$ is the 
spatially-modulated rate of Raman transition. $V_2=2\hbar U_0\cos^2(kx)|\alpha|^2$ is a standing 
wave trapping potential for atoms in state $|c\rangle$ which has been built by the cavity mode. 
The minima of $V_2$ traps the atoms in $|c\rangle$ at $x_j= j\lambda/2$ where $j$ is an integer 
and $\lambda$ is the wavelength of the cavity mode. In return, atoms in state $|c\rangle$ cause 
a shift in the cavity resonance due to their spatial overlap with the cavity mode by 
$-2U_0\int dx~|\psi_c|^2\cos^2(kx)$.

Up to this point we have included the effect of an extra parallel pump in the dynamics 
of the system. However, in Sec.\,\ref{PT} we study the phase transition without this 
parallel laser pump and later in Sec.\,\ref{practical}, where we consider practical 
ways of synthesizing a sharper and robust lattice, we will address its effect.

\section{Phase transition and formation of spin lattice}
\label{PT}
In an optical cavity, superradiance is identified by an abrupt increase in the number of 
cavity photons $n=|\alpha|^2$ and Raman transition is monitored with the total magnetization 
$Z=\int{dx~Z(x,t)}$, where $Z(x,t)=(|\psi_b(x,t)|^2-|\psi_c(x,t)|^2)/N$ is the magnetization 
density which is normalized by the total number of atoms $N=\int dx(|\psi_b(x,t)|^2+|\psi_c(x,t)|^2)$.
The total magnetization can have extremum values $1$ and $-1$ when all atoms are in mode 
$|b\rangle$ or $|c\rangle$, respectively. Therefore, if initially all atoms are in hyperfine 
state $|b\rangle$, the Raman superradiance is identified by a sudden increase in the number 
of cavity photons accompanied by an abrupt decrease in the value of total magnetization $Z$.

Dicke phase transition, in single mode condensates, takes place between two different momentum 
states of the condensate atoms which leads to density grating and is identified by an order 
parameter which measures the overlap of density distribution and cavity mode profile. In our 
system though, we will show that the density grating happens only for the atoms in hyperfine 
state $|c\rangle$ and therefore a polarization (magnetization) grating will occur. This 
translational symmetry breaking in magnetization density $Z(x,t)$ is also identified by an order 
parameter which we will introduce later in Sec.~\ref{etac}.

\begin{figure}[!t]
\begin{center}
\includegraphics[width=0.3\textwidth,angle=0]{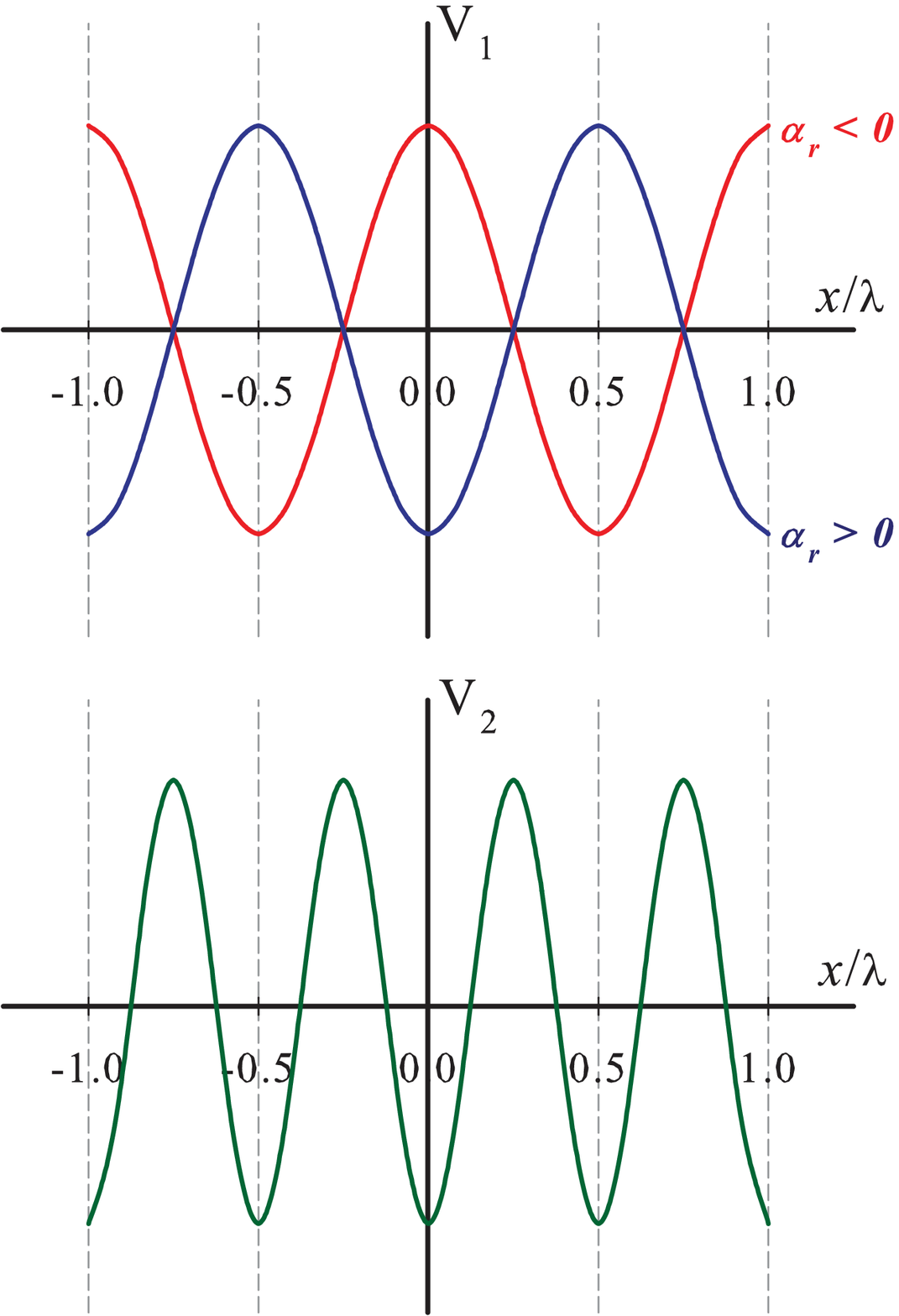}~~~~
\includegraphics[width=0.45\textwidth,angle=0]{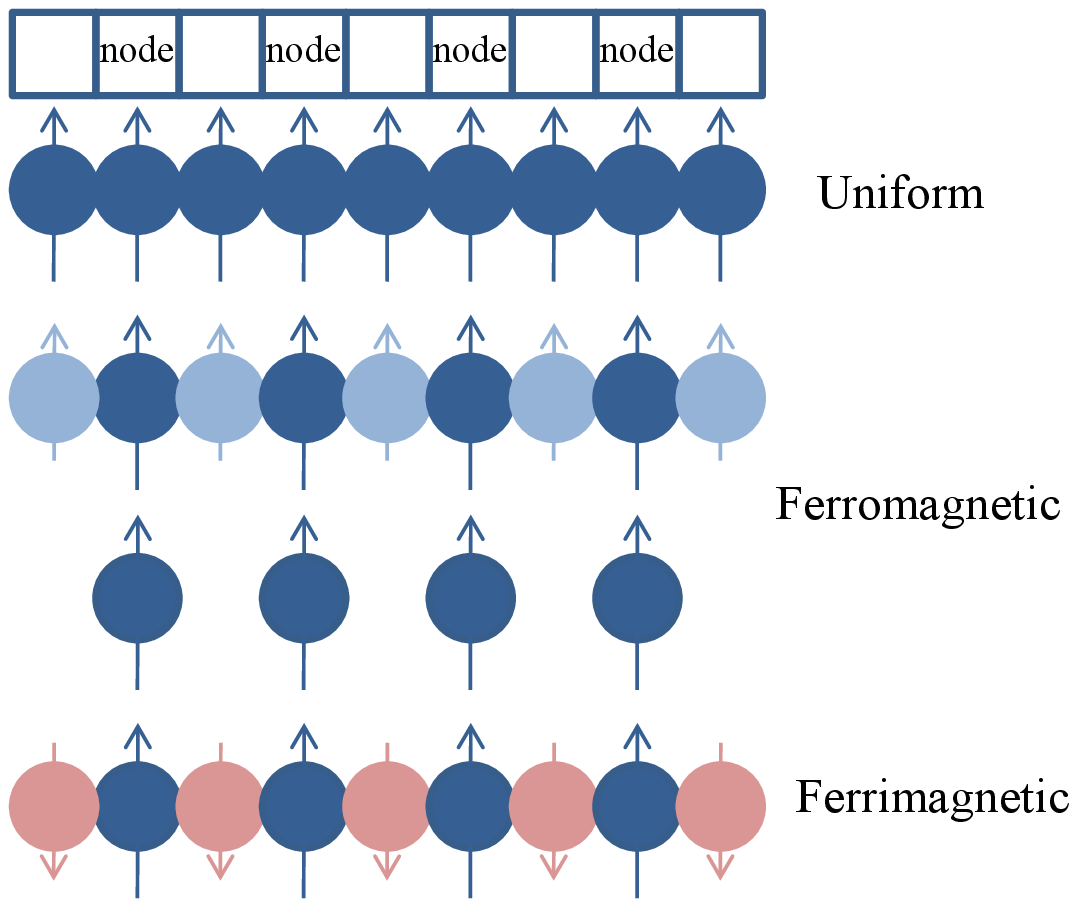}
\caption{\label{fig:potentials} (Color online) 
Left: Schematic drawing of potentials $V_1$ and $V_2$ as functions of $x$ when 
$\Delta_0$, and consequently $U_0$ and $\eta$, are chosen to be negative. $x$ 
is scaled by the wavelength of the cavity mode $\lambda$. $V_1$ defines the rate 
of Raman transition and $V_2$ is a trapping potential for atoms in hyperfine sate 
$|c\rangle$. Right: While the atoms on the nodes of $\cos(kx)$ are protected from 
Raman scattering, depending on the rate of transition $V_1$ on the antinodes, 
formation of ferromagnetic or ferrimagnetic lattices is possible. Blue circles 
with upward arrows represent positive magnetization and red circles with downward 
arrows show negative magnetization. Lighter (darker) colors and shorter (longer) 
arrows represent smaller (larger) magnetization.}
\end{center}
\end{figure}

Since the potential $V_1$ defines the rate of Raman scattering and transition of 
atoms between the two hyperfine states, the atoms on the antinodes of the $\cos(kx)$ 
are highly affected by Raman scattering (Fig.~\ref{fig:potentials}) while 
those which are on the nodes are protected. On the other hand if we choose $\Delta_0$ 
(and consequently $U_0$) to be negative, then the minima of the trapping potential $V_2$ 
will coincide with the antinodes of $V_1$. This way, overlap of different hyperfine 
spin states which acts as an atomic polarization grating is enhanced around the antinodes 
and stimulates even more Raman scattering that completes a self-consistency loop for 
a self-organization process. Accumulation of the atoms in $|c\rangle$ around the antinodes 
and protection of atoms in $|b\rangle$ from Raman scattering around the nodes of the 
cavity mode result in a spatial distillation of magnetization, manifested as a self-organized 
magnetic lattice with a lattice constant $\lambda/2$. In other words, the magnetization on the 
nodes of cavity mode will always be positive while, depending on the rate of Raman transition, 
antinodes can have positive, zero or negative magnetization (Fig.~\ref{fig:potentials}) 
and therefore synthesis of a ferromagnetic or ferrimagnetic lattice would be possible.

On the right side of Fig.~\ref{fig:potentials}, the first row from top, with dark blue circles, 
represents the uniform condensate of atoms all in state $|b\rangle$ which is the initial setup 
of the system.
Assuming that all atoms are initially in the state $|b\rangle$ practically means that the condensate 
is kept in such a low temperature that transition to higher state $|c\rangle$ due to inelastic atomic 
collisions is energetically forbidden. As an example, $^{87}$Rb condensate can have a typical temperature 
of the order of $200$ nK or less \cite{burt_coherence_1997,anderson_observation_1995}. Therefore, in 
such a condensate, transition between hyperfine states due to inelastic atomic collisions is highly 
suppressed if those states are energetically apart by $13$ KHz or more.
Second row in Fig.~\ref{fig:potentials} shows the case where less than half of 
the atoms on the antinodes are Raman scattered from $|b\rangle$ to $|c\rangle$ 
giving rise to a smaller, but still positive, value of magnetization. If fifty percent or more of 
atoms are scattered from $|b\rangle$ to $|c\rangle$, then magnetization on antinodes would become 
zero (third row) or negative (last row) resulting in formation of ferromagnetic or ferrimagnetic 
lattices.

\subsection{Critical value of pump strength for phase transition}
\label{etac}
In order to analytically calculate the critical strength of transverse pump $\eta_c$ 
for which the system undergoes superradiance and
self-organization, we first study the steady-state properties of the 
system. We assume that in the steady-state $\dot{\alpha}=0$ 
and therefore, by introducing $\theta=\int{dx~\psi_c^*(x,t)\cos(kx)\psi_b(x,t)}$, 
$\beta=\int{dx~\psi_c^*(x,t)\cos^2(kx)\psi_c(x,t)}$ and $\bar{\delta_c}=\delta_c-2U_0\beta$, 
steady cavity field can be expressed as
\begin{eqnarray}
\alpha=\frac{2\theta_r\eta}{i\kappa+\bar{\delta_c}},
\label{eq:steady-alpha}
\end{eqnarray}
with $\theta_r$ being the real part of $\theta$. As we mentioned earlier, $-2U_0\beta$ 
is the shift in the cavity mode frequency caused by atoms in state $|c\rangle$ while 
$\beta$ measures the bunching of atoms in state $|c\rangle$ inside the minima of the 
trapping potential $V_2$. The parameter $\theta$ shows the overlap of the cavity mode 
function $\cos(kx)$ with the spin polarization grating and can be considered as an 
order parameter for self-organization of magnetization. We emphasize that, in contrast 
to the case of single component BEC in optical cavity \cite{nagy_self-organization_2008}, 
$V_1$ here is the Raman transition rate and not a trapping potential. In a single component 
BEC system, different signs of the order parameter lead to two different lattice structures 
after translational symmetry breaking, when atoms are localized around the even ($kx=2n\pi$) 
or odd ($kx=(2n+1)\pi$) antinodes of the field. In our system,
breaking of the $Z_2$ symmetry also happens but it is not manifested by 
the appearance of different lattice structures. Both even and odd antinode locations are 
sites for Raman interactions that lead to the same lattice but with different magnetic character 
depending on the strength of the Raman coupling. 
Therefore, different signs of $\theta_r$ or $\alpha_r$ do not correspond to different (even and odd) 
lattice structures and a nonzero value of order parameter $\theta$ is sufficient to indicate the phase 
transition.

Regarding the wavefunctions of the condensate, in steady state, we assume that 
they can be written in the form $\psi_b(x,t)=\psi_b(x)\exp(-i\mu_bt/\hbar)$ and 
$\psi_c(x,t)=\psi_c(x)\exp(-i\mu_ct/\hbar)$, with $\mu$ being the chemical potential, 
then the dynamical equations (\ref{eq:eom-psib}) and (\ref{eq:eom-psic}) in the 
absence of external trap potentials will become
\begin{eqnarray}
\mu_b\psi_b(x)&=&
\left(
-\frac{\hbar^2}{2m}\frac{\partial^2}{\partial x^2}
+\frac{2\hbar h_0^2}{\Delta_0}
+u_{bb}|\psi_b|^2 + u_{bc}|\psi_c|^2
\right)\psi_b(x)\nonumber\\
&+&V_1\psi_c(x)e^{-\frac{i}{\hbar}\Delta\mu t}
\label{eq:steady-psib}
\end{eqnarray}
\begin{eqnarray}
\mu_c\psi_c(x)&=&
\left(
-\frac{\hbar^2}{2m}\frac{\partial^2}{\partial x^2}
+\hbar\omega_{bc}+V_2+u_{cc}|\psi_c|^2 + u_{bc}|\psi_b|^2
\right)\psi_c(x)\nonumber\\
&+&V_1\psi_b(x)e^{\frac{i}{\hbar}\Delta\mu t},
\label{eq:steady-psic}
\end{eqnarray}
where $\Delta\mu=\mu_c-\mu_b$.

For a system in which all atoms are initially in state $|b\rangle$ and are homogeneously 
distributed, the initial wavefunctions are $\psi_b(x)=\sqrt{N/L}$ and $\psi_c(x)=0$. 
Substituting these initial conditions into (\ref{eq:steady-alpha})-(\ref{eq:steady-psic}) 
results in $\alpha=0$, $\mu_b=2\hbar h_0^2/\Delta_0+u_{bb}N/L$ and 
$2\sqrt{N/L}\alpha_r\eta\cos(kx)\exp(i\Delta\mu t/\hbar)=0$. The latter is satisfied because with 
the choice of initial conditions, $\theta$ and consequently $\alpha$ are zero and it means 
that $\psi_c=0$ is a stable solution of equations of motion as long as cavity field is zero. 
Therefore to destabilize $\psi_c$, one needs to have nonzero cavity field which, in the simplest 
case, can be achieved by adding a perturbation term with $\cos(kx)$ modulation to the stable 
$\psi_c(x,0)$. Therefore the perturbed system will be defined with $\psi_b(x,t)=\sqrt{N/L}$, 
$\psi_c(x,t)=\sqrt{N/L}\epsilon\cos(kx)$ and $\alpha=N\epsilon\eta/(i\kappa+\delta_c)$. 
If this fluctuation in $\psi_c$ survives and grows, as a consequence, the order parameter 
$\theta$ and cavity field $\alpha$ will grow as well. A larger cavity field, in return, will 
advance the rate of Raman transition and will deepen the trapping potential for atoms in $|c\rangle$. 
These will lead to an even larger order parameter and, as a result of this positive feedback loop, 
superradiance and phase transition, which are characterized by an abrupt change in cavity photon 
number and magnetization, will take place. To calculate the critical value of pump strength for 
which the transition occurs, we evolve the system one step of imaginary time, starting from the 
perturbed state:
\begin{eqnarray}
\frac{\Delta\psi_b}{\Delta\tau}&=&
-\left(\frac{2\eta^2}{U_0}+\frac{N u_{bb}}{\hbar L}\right)\sqrt{N/L}
\label{eq:decay-psib}\\
\frac{\Delta\psi_c}{\Delta\tau}&=&
-\left(\omega_r+\omega_{bc}+\frac{N u_{bc}}{\hbar L}
+\frac{2N\eta^2\delta_c}{\kappa^2+\delta_c^2}
\right)
\,\sqrt{N/L}\epsilon\cos(kx),
\label{eq:decay-psic}
\end{eqnarray}
where $\tau=it$ is the imaginary time and we have used $h_0^2/\Delta_0=\eta^2/U_0$ in the 
first term on the right-hand-side of (\ref{eq:decay-psib}) in order to show the 
$\eta$-dependence of decay rates more clearly.

According to (\ref{eq:decay-psib}), $\psi_b(x,t)$ exhibits an expected decay with a rate 
equal to $\mu_b/\hbar=2h_0^2/\Delta_0+N u_{bb}/(\hbar L)$. However, situation for $\psi_c(x,t)$ 
depends on the sign of the perturbation term decay rate (terms inside the parentheses in 
(\ref{eq:decay-psic})). With the positive cavity-pump detuning $\delta_c$ this decay rate 
is always positive and perturbation will not survive. However if $\delta_c$ is negative 
then this decay rate would be negative as well if 
\begin{eqnarray}
\label{eq:etac1}
N\eta^2>
\left(\omega_r+\omega_{bc}+\frac{N u_{bc}}{\hbar L}\right)\frac{(\kappa^2+\delta_c^2)}{2|\delta_c|}
\end{eqnarray}
and therefore, we find for critical transverse pump strength $\eta_c$
\begin{eqnarray}
\label{eq:etac2}
\sqrt{N}|\eta_c|=
\sqrt{\left(\omega_r+\omega_{bc}+\frac{Nu_{bc}}{\hbar L}\right)\frac{(\kappa^2+\delta_c^2)}{2|\delta_c|}}.
\end{eqnarray}
One should notice that $\eta_c$ in (\ref{eq:etac2}) does not depend on $U_0$ 
because in our system, the effective trapping potential $V_2$ is created only for atoms in mode 
$|c\rangle$ and therefore, in return, the phase shift of the cavity mode resonance depends on the 
number of atoms in mode $|c\rangle$ which is initially negligible within the first order perturbation.
For $\omega_{bc}=\omega_r$, $N=48\times 10^3$, $L=2~\lambda$, $\kappa=400~\omega_r$ 
and $\delta_c=-4800~\omega_r$ \cite{baumann_dicke_2010}, we find $|\eta_c|\approx 2.16~\omega_r$ 
if the atom-atom interaction strength $u_{bc}\approx 3.8\times10^{-3}~\lambda\omega_r$. 

\subsection{Numerical results}
\label{numerics}
In this section the results of the numerical solution of the dynamical equations 
(\ref{eq:eom-psib}), (\ref{eq:eom-psic}) and (\ref{eq:eom-alpha}) will be presented. 
Using second order split step method, and assuming all atoms are initially in the 
hyperfine spin state $|b\rangle$, the mean-field equations are solved numerically to 
monitor the dynamics of the system.

\begin{figure}[!t]
\begin{center}
\includegraphics[width=0.9\textwidth]{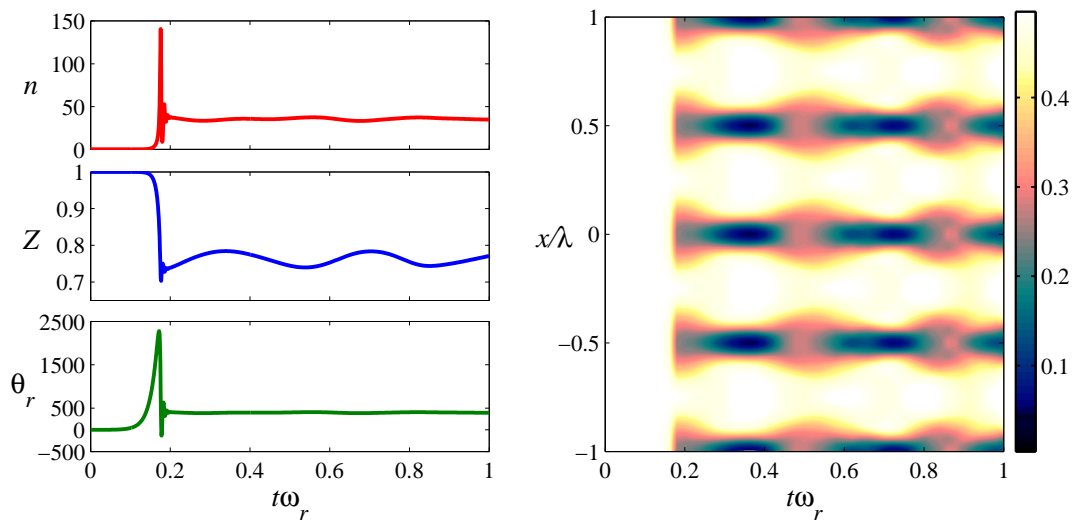}
\caption{\label{fig:eta3} (Color online) Left: Dynamics of the cavity photon number 
$n$, total magnetization $Z$ and the real part of order parameter $\theta_r$. Right: 
spatial and dynamical behavior of magnetization density $Z(x,t)$ of the same system. 
Here, system is subject to a transverse pump with fixed strength $\eta=-3~\omega_r$ 
and other parameters, in units of $\omega_r$, are $\omega_{bc}=1$, $\kappa=400$, 
$U_0=-0.5$, $\delta_c=-4800$, $\Delta_0=-4\times 10^6$ and $N=48\times 10^3$.}
\end{center}
\end{figure}

Fig.~\ref{fig:eta3} demonstrates the dynamics of the cavity photon number $n=|\alpha|^2$, 
total magnetization $Z$, the order parameter $\theta_r$ (all on the left panels) as well 
as the spatio-temporal behavior of magnetization density $Z(x,t)$ (on the right) when the 
transverse pump has the strength $\eta=-3~\omega_r$. We have considered a cavity with 
wavelength $\lambda=800~\text{nm}$ \cite{baumann_dicke_2010} which for Rubidium atoms gives 
a recoil frequency of $\omega_r\sim 20$ KHz. Other parameters used for this simulation, in 
units of $\omega_r$, are $\omega_{bc}=1,~\kappa=400$, $\delta_c=-4800$ \cite{baumann_dicke_2010}, 
$U_0=-0.5$, and $\Delta_0=-4\times 10^6$. We have considered a condensate of $N=48\times 10^3$, 
atom-atom interaction strengths $u_{bc}=u_{cb}\approx 3.8\times10^{-3}~\lambda\omega_r$ and 
$u_{bb}=u_{cc}\approx 4.5\times10^{-3}~\lambda\omega_r$. One can see in Fig.~\ref{fig:eta3} 
that superradiance and phase transition take place at $t\sim 0.2~\omega_r$ after the system 
is pumped with the transverse laser with strength $\eta=-3~\omega_r$. 
While cavity photon number, total magnetization and order parameter reach slowly-oscillating 
steady states, a ferromagnetic lattice of magnetization is formed. Since atoms initially were 
in state $|b\rangle$, total magnetization $Z$ is equal to one before the phase transition. On 
the other hand atoms are initially distributed in an area with length $L=2\lambda$ homogeneously 
which gives rise to magnetization $Z(x,t)=0.5$ throughout the condensate. In this case after 
the transition less than (but very close to) fifty percent of atoms on the antinodes of the 
potential $V_1$ are scattered to state $|c\rangle$, causing a very small positive value of 
magnetization around the antinodes. However, the atoms on the nodes are almost untouched as 
expected.

\begin{figure}[!t]
\begin{center}
\includegraphics[width=0.9\textwidth]{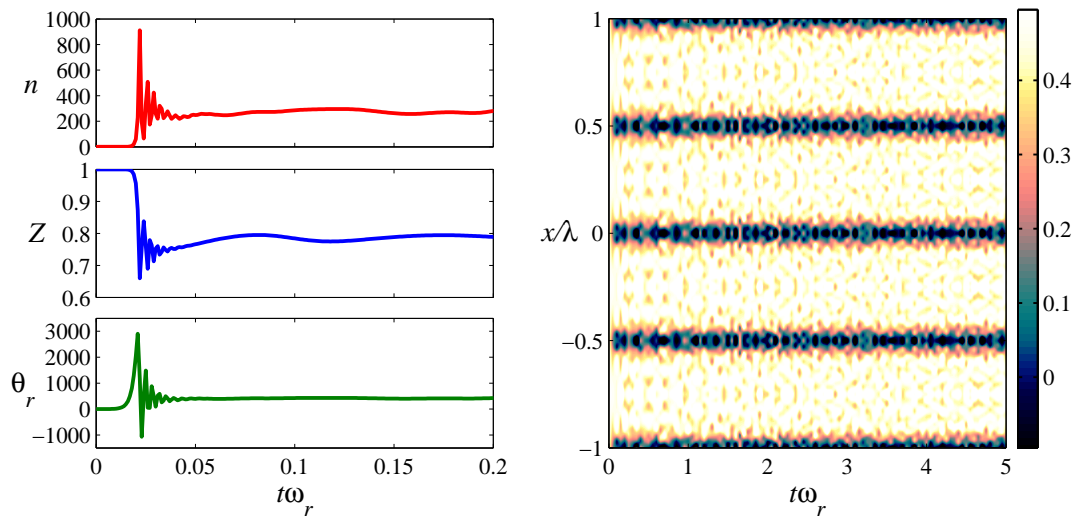}
\caption{\label{fig:eta8} (Color online) Left: Dynamics of the cavity photon number 
$n$, total magnetization $Z$ and the real part of order parameter $\theta_r$. Left: 
spatial and dynamical behavior of magnetization density $Z(x,t)$ of the same system 
is shown in a wider range of time to give a better view of the synthesized lattice. 
In this case, system is subject to a transverse pump with fixed strength $\eta=-8~\omega_r$ 
and other parameters are the same as those in Fig.~\ref{fig:eta3}}
\end{center}
\end{figure}

Since the scattering rate $V_1$ and cavity field $\alpha$ are proportional to the transverse 
pump strength $\eta$, one would expect a higher percentage of atoms around the antinodes of 
$V_1$ being scattered to $|c\rangle$ by simply using a larger $\eta$. Fig.~\ref{fig:eta8} shows 
the dynamics of the system with $\eta=-8~\omega_r$ where a ferrimagnetic lattice of magnetization 
is created due to transition of more than fifty percent of atoms on antinodes from state $|b\rangle$ 
to $|c\rangle$. Apart from $\eta$, all other parameters are similar to those used in Fig.~\ref{fig:eta3} 
and spatio-temporal behavior of magnetization $Z(x,t)$ is shown for a longer time in order to 
present a clearer view of the spin lattice.

Although the perturbation method in Sec.\,\ref{etac} predicts a phase transition for 
the transverse pump with strength $2.16~\omega_r$ or above, numerical solution leads to 
a phase transition with values of pump strength smaller than the value predicted by the 
perturbation method. In numerical method, no phase transition occurs with 
$\eta\leq1.85~\omega_r$.
%
\section{Practical synthesis of a robust spin lattice}
\label{practical}
In this section we address some practical issues which might be helpful in the implementation 
of a sharp and robust spin-lattice. First of all we remind that in the last section the strength 
of the transverse pump was assumed to be constant. However, considering a time-dependent pump 
is more practical. In the experiments the power of the pump is usually ramped up in time such 
that it is initially zero and increases gradually. In numerics, using a ramped-up pump delays 
the time of transition because during early stages the system is subject to a laser with smaller 
values of strength. 
This would give more control on the system at the time of transition. Moreover, one would think 
of having a robust lattice without the need of an all-time-on laser field. In other words, it 
would be desirable to turn off the laser pump after synthesis of the spin lattice. We will show 
numerically that it is possible to have a robust lattice even when the laser pump is switched 
off after the transition. The fact that the atoms on the nodes of $V_1$ remain untouched and 
therefore around the nodes there exist a single-component condensate while around the antinodes 
both modes are occupied is the reason of this robustness. When the pump is switched off ($\eta=0$), 
the Raman coupling terms (last terms) in (\ref{eq:eom-psib}) and (\ref{eq:eom-psic}) vanish and 
these equations are reduced to equations of motion of a two-component condensate with atom-atom 
interaction. Due to the difference between the chemical potentials of the two components, there 
will be coherent oscillations in their wavefunctions~\cite{busch_dark-bright_2001, das_sinusoidal_2009} 
and consequently in the magnetization. This is the case around the antinodes while, around the nodes, 
single-component condensate remains stable.

\begin{figure}[!t]
\begin{center}
\includegraphics[width=0.90\textwidth]{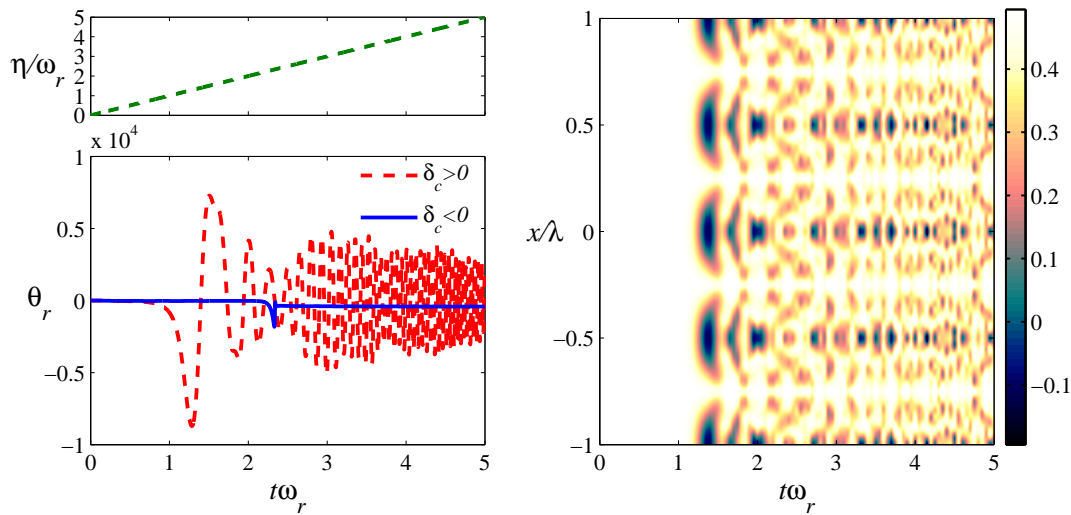}
\caption{\label{fig:ramping} (Color online) 
Left: Time dependence of strength of effective transverse laser pump $\eta$ and 
dynamics of real part of order parameter $\theta_r$, with fixed parallel pump 
strength $\eta_{\|}=1000~\omega_r$ and for positive and negative cavity-pump 
detuning $\delta_c$. Right: Spatial and dynamical behavior of magnetization density 
$Z(x,t)$ for the case of positive $\delta_c$ which shows synthesis of a sharp 
lattice around time $t\approx 1.4/\omega_r$ when $\eta\approx 1.4~\omega_r$. 
Here, in units of $\omega_r$, $\omega_{bc}=1$, $\kappa=400$, $U_0=-0.5$, $|\delta_c|=4800$, 
$\Delta_0=-4\times 10^6$ and $N=48\times 10^3$.}
\end{center}
\end{figure}

Another point to be considered is the role of an extra laser pump, parallel to the cavity axis. 
A parallel laser pump can contribute to the cavity field as is shown in (\ref{eq:eom-alpha}) 
such that the steady cavity field takes the following form
\begin{eqnarray}
\alpha=\frac{2\theta_r\eta+i\eta_{\|}}{i\kappa+\bar{\delta_c}}.
\label{eq:steady-alpha2}
\end{eqnarray}
In addition to seeding the cavity field instead of relying on fluctuations to trigger the 
phase transition, a strong parallel pump can also affect the depth of the trapping potential 
$V_2$ as well as the rate of scattering $V_1$ indirectly through the cavity field. These 
latter facts would allow formation of a sharper spin lattice with smaller values of transverse 
pump strength. More interestingly, through its effect on $V_2$ and consequently the decay 
rate of fluctuations in $\psi_c$, a parallel pump can open up the possibility of formation 
of a spin lattice with both positive and negative cavity-pump detuning $\delta_c$.

To bring all the above points together, in Fig~\ref{fig:ramping} we show the real part of 
order parameter $\theta_r$ as a function of time for the two cases with positive and negative 
cavity-pump detuning $\delta_c$, while parallel pump strength is fixed to $\eta_{\|}=1000~\omega_r$ 
and transverse pump strength is ramped up from zero at $t=0$ to $\eta=-5~\omega_r$ at time 
$t=5/\omega_r$. In the case with positive $\delta_c$, cavity photon number $n$ and magnetization 
$Z$ exhibit oscillatory behavior similar to $\theta_r$ after the transition. For the case with 
negative $\delta_c$, $Z$ saturates to a stationary value while $n$ increases due to the 
increase in the pump strength. 

In both cases, with positive or negative $\delta_c$, the order parameter is initially zero 
as a sign of a homogeneous condensate. Then, when a critical value of transverse pump strength 
$\eta$ is reached, Raman superradiance takes place and simultaneously a polarization grating 
happens due to accumulation of atoms in state $|c\rangle$ on the antinodes of the cavity mode 
function. As a consequence of the translational symmetry breaking, the value of $\theta_r$ becomes 
nonzero. Since any change in the value of the order parameter is a sign of the change in the 
value or distribution of magnetization around the antinodes of the cavity mode, by looking at 
the oscillations of $\theta_r$ in Fig.~\ref{fig:ramping} for positive $\delta_c$, one would 
expect oscillations in the magnetization density around the antinodes for this case. On the 
right panel of Fig.~\ref{fig:ramping} the spatio-temporal behavior of magnetization density 
$Z(x,t)$ of the case with positive $\delta_c$, is also shown which demonstrates expected oscillations. 
The case with negative $\delta_c$ exhibits a stable lattice structure, however, the lattice 
is never as well-defined as the one with positive $\delta_c$ at $t\sim 1.4/\omega_r$. The value 
of the order parameter on left side of Fig.~\ref{fig:ramping} is a clear sign of this fact.   

\begin{figure}[!t]
\begin{center}
\includegraphics[width=0.90\textwidth]{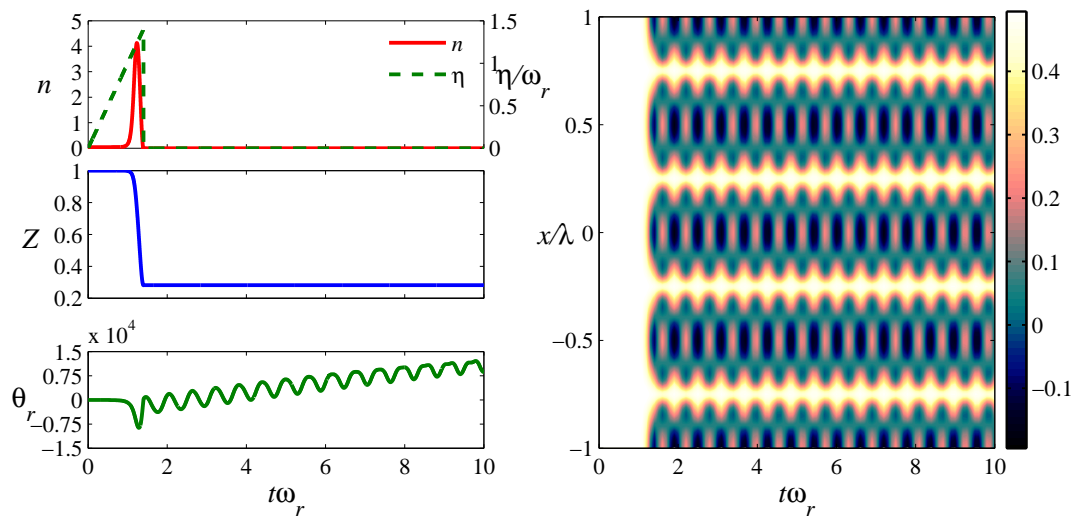}
\caption{\label{fig:turningoff} (Color online) 
Left: Time dependence of strength of effective transverse laser pump $\eta$ 
(dashed line) and dynamics of the cavity photon number $n$, total magnetization 
$Z$ and real part of order parameter $\theta_r$. Right: Spatial and dynamical 
behavior of magnetization density $Z(x,t)$. In this case parallel pump strength 
has been fixed to $\eta_{\|}=1000~\omega_r$ while transverse pump strength has 
been ramped up from zero at $t=0$ to $\eta=-1.4~\omega_r$ at $t=1.4/\omega_r$ 
and then both of pumps are turned off when the spin lattice has taken a well-defined 
shape. Other parameters are the same as those in Fig.~\ref{fig:ramping} 
with positive $\delta_c$.}
\end{center}
\end{figure}

Although magnetization density of the case with positive $\delta_c$ oscillates in time, it 
clearly demonstrates synthesis of a very sharp lattice around $t\sim 1.4/\omega_r$ when 
$\eta\sim -1.4~\omega_r$. From the right panel of Fig.~\ref{fig:ramping} one can observe 
that, as the pump is ramped up, the magnetized domains are not well isolated from the 
de-magnetized ones and there is no robust spin lattice structure for this case. This problem 
can be avoided by turning both laser pumps off when the lattice has taken the desired 
shape. Fig.~\ref{fig:turningoff} shows dynamics of cavity photon number $n$, total 
magnetization $Z$, order parameter $\theta_r$, as well as magnetization density $Z(x,t)$ 
for the case with positive $\delta_c$ and fixed parallel pump $\eta_{\|}=1000~\omega_r$, 
when both pumps are abruptly turned off at time $t=1.4/\omega_r$. 
At this point, due to the lack of Raman transition, magnetization $Z$ 
remains a constant and all the photons leave the cavity. 
In the absence of the laser pumps, the regions around the antinodes of the 
cavity mode exhibit oscillatory behavior as expected. However, nodes with maximum positive 
magnetization remain untouched and are very well separated from each other. Since there is 
no laser pumping the system, one would not expect a net change in the order parameter. In 
fact, the increase in $\theta_r$, observed in Fig.~\ref{fig:turningoff}, is a part of a 
very slow oscillatory behavior. In other words, $\theta_r$ exhibits some fast oscillations 
as well as slow oscillations. 
 
\section{Conclusion}
\label{conc}
We conclude that a BEC with two non-degenerate hyperfine spin components in a high-finesse 
cavity driven by a transverse pump can exhibit Raman superradiance above a critical value of 
the transverse field strength. Simultaneously, BEC undergoes a phase transition, associated both 
with the external and internal degrees of freedom, during which atoms scatter transverse laser 
field into the cavity mode and in return their hyperfine state changes. As a result, cavity 
photon number rises abruptly and at the same time there is a sudden increase in the population 
of higher hyperfine state at periodic positions exhibiting a magnetic lattice configuration. An extra 
laser pump parallel to the cavity axis can be used in order to synthesize sharper lattices and 
the lattice remains robust after turning both laser pumps off. 
Even though present analysis is in the mean-field regime, we can envision that hyperfine spins 
at different lattice sites would be entangled as they interact with the common cavity field, 
following the resonant entanglement of atoms in multitraps scenario \cite{yukalov_regulating_2006}. 
Availability of large number of spins per site could make the system advantageous for explorations 
of magnetic supersolid properties. In contrast to Rayleigh superradiance, Raman superradiance can 
be used as a source of entangled photon-spin pairs. 
Application of cold atoms in optical lattices for quantum information purposes has been a developing 
field of theoretical and experimental studies \cite{braunstein_scalable_2001, jaksch_the_2005,
treutlein_quantum_2006}. In addition to optical lattices, spin systems are commonly considered for 
quantum information bits (qubits) and associated quantum information processing. Our treatment brings 
optical lattices and spin lattices together in a compact and controllable cavity-QED environment.
Synthesis and probing robust spin lattice models, with fast superradiance induced phase transition and 
self-organization properties promise unique opportunities for quantum information applications as well 
as monitoring phase transitions and spin correlations with Raman scheme. 
Long range spin-spin interactions induced by the cavity field can be utilized on optical spin lattice 
created in the cavity. Large spin values at the sites together with the coherence from the underlying 
condensate can be useful for quantum memory as well as information processing. Moreover, leaking 
photons from the cavity and external drives can be used for non-destructive probing and accessing 
the system. We hope our work will stimulate further research in this direction.
%
\ack
This work was supported by T\"UB\.{I}TAK (Grant No. 109T267, 112T176 and 112T974) and TUBA.


\section*{References}
\begin{thebibliography}{}

\bibitem{mekhov_quantum_2012}
	I. B. Mekhov and H. Ritsch, 
	J. Phys. B: At. Mol. Opt. Phys.~{\bf 45}, 102001 (2012).
\bibitem{ritsch_cold_2013}
	H. Ritsch, P. Domokos, F. Brennecke, and T. Esslinger, 
	Rev. Mod. Phys.~{\bf 85}, 553 (2013).
\bibitem{domokos_collective_2002}
	P. Domokos and H. Ritsch, 
	\prl~{\bf 89}, 253003 (2002).
\bibitem{black_observation_2003}
	A. T. Black, H. W. Chan, and V. Vuleti{\'c}, 
	\prl~{\bf 91}, 203001 (2003).
\bibitem{nagy_self-organization_2008}
	D. Nagy, G. Szirmai, and P. Domokos,
	Eur. Phys. J. D~{\bf 48}, 127 (2008).
\bibitem{konya_multimode_2011}
	G. K\'onya, G. Szirmai, and P. Domokos, 
	Eur. Phys. J. D~{\bf 65}, 33 (2011).
\bibitem{bux_cavity-controlled_2011}
	S. Bux, C. Gnahm, R. A. W. Maier, C. Zimmermann, and Ph. W. Courteille, 
	\prl~{\bf 106}, 203601 (2011).	
\bibitem{horak_coherent_2000}
	P. Horak, S. M. Barnett, and H. Ritsch, 
	\pra~{\bf 61}, 033609 (2000).	
\bibitem{brennecke_cavity_2007}
	F. Brennecke, T. Donner, S. Ritter, T. Bourdel, M. K\"ohl, and T. Esslinger, 
	\nt~{\bf 450}, 268 (2007).
\bibitem{colombe_strong_2007}
  Y. Colombe, T. Steinmetz, G. Dubois, F. Linke, D. Hunger, and J. Reichel, 
 \nt~{\bf 450}, 272 (2007).
\bibitem{slama_superradiant_2007}
	S. Slama, S. Bux, G. Krenz, C. Zimmermann, and Ph. W. Courteille, 
	\prl~{\bf 98}, 053603 (2007).	
\bibitem{wolke_cavity_2012}
	M. Wolke, J. Klinner, H. Ke{\ss}ler, and A. Hemmerich, 
	\sci~{\bf 337}, 75 (2012).	
\bibitem{elsasser_optical_2004}
	Th. Els\"asser, B. Nagorny, and A. Hemmerich, 
	\pra~{\bf 69}, 033403 (2004). 
\bibitem{gupta_cavity_2007}
	S. Gupta, K. L. Moore, K. W. Murch, and D. M. Stamper-Kurn, 
	\prl~{\bf 99}, 213601 (2007).
\bibitem{ritter_dynamical_2009}
	S. Ritter, F. Brennecke, K. Baumann, T. Donner, C. Guerlin, and T. Esslinger, 
	App. Phys. B~{\bf 95}, 213 (2009).
\bibitem{safaei_bistable_2013}	
	S. Safaei, \"O. E. M\"ustecapl{\i}o\u{g}lu, and B. Tanatar, 
	Laser Phys.~{\bf 23}, 035501 (2013) .
\bibitem{mekhov_probing_2007}
	I. B. Mekhov, C. Maschler, and H. Ritsch, 
	\nt~Physics~{\bf 3}, 319 (2007).
\bibitem{dicke_coherence_1954}
	R. H. Dicke, 
	Phys. Rev.~{\bf 93}, 99 (1954).
\bibitem{hepp_superradiant_1973}
	K. Hepp and E. H. Lieb, 
	Ann. Phys.~{\bf 76}, 360 (1973).
\bibitem{wang_phase_1973}
	Y. K. Wang and F. T. Hioe, 
	\pra~{\bf 7}, 831 (1973).
\bibitem{nagy_dicke-model_2010}
	D. Nagy, G. K\'onya, G. Szirmai, and P. Domokos, 
	\prl~{\bf 104}, 130401 (2010).
\bibitem{baumann_dicke_2010}
	K. Baumann, C. Guerlin, F. Brennecke, and T. Esslinger, 
	\nt~{\bf 464}, 1301 (2010).
\bibitem{baumann_exploring_2011}
	K. Baumann, R. Mottl, F. Brennecke, and T. Esslinger, 
	\prl~{\bf 107}, 140402 (2011).
\bibitem{bhaseen_dynamics_2012}
	M. J. Bhaseen, J. Mayoh, B. D. Simons, and J. Keeling, 
	\pra~{\bf 85}, 013817 (2012).
\bibitem{torre_keldysh_2013}
	E. G. Dalla Torre, S. Diehl, M. D. Lukin, S. Sachdev, and P. Strack, 
	\pra~{\bf 87}, 023831 (2013).
\bibitem{konya_finite_2012}
	G. K\'onya, D. Nagy, G. Szirmai, and P. Domokos, 
	\pra~{\bf 86} 013641 (2012).
\bibitem{oztop_excitations_2012} 
	B. \"Oztop, M. Bordyuh, \"O. E. M\"ustecapl{\i}o\u{g}lu, and H. E. T\"ureci, 
	New J. Phys.~{\bf 14}, 085011 (2012).	
\bibitem{nagy_critical_2011}
	D. Nagy, G. Szirmai, and P. Domokos, 
	\pra~{\bf 84} 043637 (2011).	
\bibitem{yukalov_cold_2009}
	V. I. Yukalov, 
	\lp~{\bf 19}, 1 (2009).
\bibitem{schneble_raman_2004}	
  D. Schneble, G. K. Campbell, E. W. Streed, M. Boyd, D. E. Pritchard, and W. Ketterle, 
  \pra~{\bf 69}, 041601(R) (2004).
\bibitem{yoshikawa_observation_2004}
  Y. Yoshikawa, T. Sugiura, Y. Torii, and T. Kuga, 
  \pra~{\bf 69}, 041603(R) (2004).
\bibitem{cola_theory_2004}
	Mary M. Cola and Nicola Piovella, 
	\pra~{\bf 70}, 045601 (2004).
\bibitem{uys_theory_2007}
	H. Uys and P. Meystre, 
	\pra~{\bf 75}, 033805 (2007).
\bibitem{huang_cavity-induced_2011}
	J.-S. Huang, Z.-W. Xie, and L.-F. Wei, 
	Commun. Theor. Phys.~{\bf 55}, 59 (2011).
\bibitem{strack_dicke_2011}
	P. Strack and S. Sachdev, 
	\prl~{\bf 107}, 277202 (2011).
\bibitem{zhou_cavity-mediated_2009}
	L. Zhou, H. Pu, H. Y. Ling, and W. Zhang, 
	\prl~{\bf 103}, 160403 (2009).
\bibitem{zhou_spin_2010}
	L. Zhou, H. Pu, H. Y. Ling, K. Zhang, and W. Zhang, 
	\pra~{\bf 81}, 063641 (2010).
\bibitem{dong_multistability_2011}
	Y. Dong, J. Ye, and H. Pu, 
	\pra~{\bf 83}, 031608(R) (2011).
\bibitem{grieser_selforganisation_2011}
	T. Grie{\ss}er, W. Niedenzu, and H. Ritsch, 
	New J. Phys.~{\bf 14}, 053031 (2012).
\bibitem{guo_cavity-enhanced_2009}
	L. Guo, S. Chen, B. Frigan, L. You, and Y. Zhang, 
	\pra~{\bf 79}, 013630 (2009).
\bibitem{kagan_effect_1985}
	Yu. Kagan, B. V. Svistunov, and G. V. Shlyapnikov, 
	Sov. Phys. JETP Lett. {\bf 42}, 209 (1985).
\bibitem{burt_coherence_1997}
	E. A. Burt, R. W. Ghrist, C. J. Myatt, M. J. Holland, E. A. Cornell, and C. E. Wieman, 
	\prl~{\bf 79}, 337 (1997).	
\bibitem{ketterle_coherence_1997}
	W. Ketterle and H. J. Miesner, 
	\pra~{\bf 56} 3291 (1997).
\bibitem{gangl_cavity_2002}
	M. Gangl and H. Ritsch, 
	J. Phys. B: At. Mol. Opt. Phys.~{\bf 35}, 4565 (2002).
\bibitem{anderson_observation_1995}
	M. H. Anderson, J. R. Ensher, M. R. Matthews, C. E. Wieman, and E. A. Cornell, 
	\sci~{\bf 269}, 198 (1995).
\bibitem{busch_dark-bright_2001}
	Th. Busch and J. R. Anglin, 
	\prl~{\bf 87}, 010401 (2001).
\bibitem{das_sinusoidal_2009}
	P. Das, T. S. Raju, U. Roy, and P. K. Panigrahi, 
	\pra~{\bf 79}, 015601 (2009).
\bibitem{yukalov_regulating_2006}
	V. I. Yukalov and E. P. Yukalova, 
	\pra~{\bf 73}, 022335 (2006).	
\bibitem{braunstein_scalable_2001}
	S. L. Braunstein and Hoi-Kwong Lo, 
	{\it{Scalable Quantum Computers: Paving the Way to Realization}}, Wiley-VCH (2001).
\bibitem{jaksch_the_2005}
	D. Jaksch and P. Zoller, 
	Annals of Physics~{\bf 315}, 52 (2005).
\bibitem{treutlein_quantum_2006}
	P. Treutlein, T. Steinmetz, Y. Colombe, B. Lev, P. Hommelhoff, J. Reichel, M. Greiner, 
	O. Mandel, A. Widera, T. Rom, I. Bloch, and T. W. H\"ansch, 
	Fortschr. Phys.~{\bf 54}, 702 (2006).

\end{thebibliography}
\end{document}